\documentclass[12pt]{article}

\usepackage{amsmath}
\usepackage{amssymb}
\usepackage{hyperref}
\usepackage[final]{graphicx}
\usepackage{bbm,latexsym,epsfig}

\newcommand{\bs}{\begin{subequations}}
\newcommand{\es}{\end{subequations}}
\newcommand{\be}{\begin{equation}}
\newcommand{\ee}{\end{equation}}
\newcommand{\ba}{\begin{eqnarray}}
\newcommand{\ea}{\end{eqnarray}}
\newcommand{\no}{\nonumber \\}

\newcommand{\z}{\mathbbm{Z}}
\newcommand{\zz}{\mathbbm{Z}_2}
\newcommand{\zzz}{\mathbbm{Z}_3}
\newcommand{\zzzz}{\mathbbm{Z}_4}
\newcommand{\bone}{\mathbbm{1}}
\newcommand{\three}{\mathbf{3}}

\allowdisplaybreaks

\begin{document}

\title{
\normalsize \hfill UWThPh-2013-19
\\[.5mm]
\normalsize \hfill CFTP/13-020
\\[10mm]
\LARGE Double seesaw mechanism and lepton mixing}

\author{
W.~Grimus$^{(1)}$\thanks{E-mail: \tt walter.grimus@univie.ac.at}
\ and
L.~Lavoura$^{(2)}$\thanks{E-mail: \tt balio@cftp.ist.utl.pt}
\\*[8mm]
$^{(1)} \! $
\small University of Vienna, Faculty of Physics, \\
\small Boltzmanngasse 5, A--1090 Wien, Austria
\\[3mm]
$^{(2)} \! $
\small Universidade de Lisboa, Instituto Superior T\'ecnico, CFTP, \\
\small 1049-001 Lisboa, Portugal
\\*[5mm]
}

%%%%% \date{January 15, 2014}
\date{January 28, 2014}
\maketitle

\begin{abstract}
We present a general framework for models in which
the lepton mixing matrix is the product
of the maximal mixing matrix $U_\omega$
times a matrix constrained by a well-defined $\zz$ symmetry.
Our framework relies on
neither supersymmetry
nor non-renormalizable Lagrangians
nor higher dimensions;
it relies instead on the double seesaw mechanism
and on the soft breaking of symmetries.
The framework may be used to construct models
for virtually all the lepton mixing matrices of the type mentioned above
which have been proposed in the literature.
\end{abstract}

\newpage

\section{Introduction}
With the measurement of the reactor mixing angle
$\theta_{13}$~\cite{daya-reno},  
our knowledge of the lepton mixing matrix $U$
is almost complete.\footnote{For global fits of $U$ see ref.~\cite{schwetz}.}
Only the CKM-type phase~$\delta$ is still unknown.
Because $s_{13}^2$
($s_{ij} \equiv \sin \theta_{ij}$ and 
$c_{ij} \equiv \cos \theta_{ij}$ for $i,j = 1,2,3$)
is definitely nonzero, 
strict tri-bimaximal mixing (TBM)~\cite{HPS} is ruled out. 
However,
it is still viable to relax TBM in such a way that
either the first column ($u_1$) or the second column ($u_2$)
of $U = \left( u_1,\, u_2,\, u_3 \right)$ coincides with its form in TBM.
Let TM$_1$ and TM$_2$,
respectively,
denote these two possibilities~\cite{albright}.
In a suitable phase convention,
one has
\bs
\ba
\label{TM1}
\mbox{TM}_1: & &
u_1 = \frac{1}{\sqrt{6}} \left( \begin{array}{c} 2 \\ -1 \\ -1 
\end{array} \right),
\\*[2mm]
\label{TM2}
\mbox{TM}_2: & &
u_2 = \frac{1}{\sqrt{3}} \left( \begin{array}{c} 1 \\ 1 \\ 1 
\end{array} \right).
\ea
\es

On the other hand,
many authors~\cite{residual} have recently pursued an approach
in which $U$ is either partially or completely determined
by distinct symmetries in the charged-lepton mass matrix $M_\ell$
and in the light-neutrino Majorana mass matrix
$M_\nu$. Those distinct symmetries are conceived as remnants
of the full flavour symmetry group of the Lagrangian.
In particular,
in ref.~\cite{lindner} three candidates for a completely determined $U$
have been found in this way.
In that approach,
$U$ can be written,
in the weak basis in which the models are formulated,
as the product
\be
\label{UV}
U = U_\omega V,
\ee
where
\be
\label{Uomega}
U_\omega \equiv \frac{1}{\sqrt{3}} \left( \begin{array}{ccc}
1 & 1 & 1 \\ 1 & \omega & \omega^2 \\ 1 & \omega^2 & \omega
\end{array} \right)
\ee
($\omega \equiv \exp{\left( 2 \pi i / 3 \right)}$)
is a symmetric unitary matrix,
which diagonalizes $M_\ell$ in that weak basis,
while $V$ is the unitary matrix that diagonalizes $M_\nu$.
In this approach the matrix $V$ is constrained 
by either one or two well-defined $\zz$ symmetries,
leading to the determination of either one column or all the columns of $U$,
respectively. 

Unfortunately,
it is not easy to implement the approach of the previous paragraph
in the context of well-defined,
self-contained models.
The same happens if one tries to implement TM$_1$
in such models.\footnote{For such implementations
see for instance refs.~\cite{tm1,varzielas}.}
One usually has recourse to supersymmetry,
non-renormalizable Lagrangians,
and additional superfields
(`familons' and `driving fields'),
or else to theories with extra dimensions; 
the resulting models tend to be complicated and unaesthetic.

In this paper we present a framework which allows one to realize,
in a technically natural way,
mixing matrices with either TM$_1$,
TM$_2$,
or virtually any of the viable mixing matrices
found in refs.~\cite{residual,lindner}.
Our framework relies on the double seesaw mechanism
and on a soft flavour symmetry breaking
in the Majorana mass matrix of right-handed neutrino
singlets;
it involves neither of the technical complications
mentioned in the previous paragraph.
On the other hand,
our framework rests on two assumptions about the vacuum,
which is characterized by two vastly different scales:
the vacuum must preserve a symmetry of the Lagrangian at the high scale
and break another symmetry of the Lagrangian at the low scale.
Actually,
it should be possible to implement these two assumptions
by choosing suitable ranges for the parameters in the scalar potential. 
Unfortunately,
we have been unable to prove this in all generality and beyond doubt.

Let us first discuss the double seesaw mechanism~\cite{barr}.
It is based on the existence of right-handed neutrinos
(gauge singlets)
beyond the three usual ones found in
the standard seesaw mechanism~\cite{seesaw}.\footnote{Additional
right-handed neutrinos were also used
in our previous models of ref.~\cite{only},
but in those models we did not use the double seesaw mechanism.}
Specifically,
the models presented in this paper have six right-handed neutrinos;
let $\nu_{jR}$ and $N_{jR}$ ($j=1,2,3$) denote them.
The effective mass Lagrangian of the neutrinos
has the usual form
\be
\label{numass}
\mathcal{L}_{\nu\, \mathrm{mass}} = 
- \left( \bar \nu_R,\, \bar N_R \right) M_D \nu_L
- \frac{1}{2}
\left( \bar \nu_R,\,  \bar N_R \right) M_R\, C 
\left( \begin{array}{c} \bar \nu_R^T \\[1mm] \bar N_R^T 
\end{array} \right)
+ \mbox{H.c.}
\ee
In eq.~(\ref{numass}),
$\nu_L$ denotes the column vector of the standard three left-handed neutrinos
belonging to doublets of weak isospin.
The matrix $M_D$ is $6 \times 3$
while $M_R$ is $6 \times 6$ and symmetric.
They act in family space;
the charge-conjugation matrix $C$ acts in Dirac space.

The $\nu_{jR}$ and $N_{jR}$ are distinguished by two features:
firstly,
the $\nu_{jR}$ have Yukawa couplings
to the leptonic weak-isospin doublets
but the $N_{jR}$ do not;
secondly,
there are no Majorana mass terms among the $\nu_{jR}$.
(Evidently,
both these features are in each specific model
enforced by well-defined symmetries of the model.)
Thus,
\bs
\label{zero}
\ba
\label{md}
M_D &=& \left( \begin{array}{c} Y \\ 0 \end{array} \right),
\\
\label{mr}
M_R &=& \left( \begin{array}{cc} 0 & X \\ X^T & M \end{array} \right),
\ea
\es
where $X$,
$Y$,
$M$,
and the null matrix are all $3 \times 3$ matrices
($M$ is furthermore symmetric).\footnote{The presence
of both $X$ and $M$ in eq.~(\ref{mr})
does not imply that the $\nu_R$ and the $N_R$
transform in the same way under the symmetries of the model.
Indeed,
either one or both those matrices---and possibly also $Y$
in eq.~(\ref{md})---result from the spontaneous breaking
of flavour symmetries of the model.
A double-seesaw model must therefore comprehend many scalars.}
We assume that the mass scale $m_X$ inherent in $X$
is much larger than the mass scale $m_Y$ of $Y$.\footnote{The matrix $X$
is moreover assumed to be non-singular.}
Then the standard seesaw formula
\be
M_\nu = - M_D^T M_R^{-1} M_D
\label{standardseesaw}
\ee
applies.
Since
\be
M_R^{-1} =
\left( \begin{array}{cc}
- {X^T}^{-1} M X^{-1} & {X^T}^{-1} \\*[1mm] X^{-1} & 0
\end{array} \right),
\ee
one has
\be
M_\nu = Y^T {X^T}^{-1} M X^{-1} Y.
\label{mnu}
\ee
One furthermore assumes that the mass scale $m_{\rm soft}$ of $M$
is much smaller than the Fermi scale;
hence a double suppression of the neutrino masses---by $m_Y / m_X$
and by $m_{\rm soft} / m_{\rm Fermi}$---that has been dubbed
`double seesaw mechanism'.
The smallness of $m_{\rm soft}$ is usually explained in a technically natural
way by the additional
(fundamental or accidental)
lepton-number symmetry that exists when $M=0$.
Indeed,
in that limit the $\nu_L$ and $\nu_R$ have conserved lepton number $+1$
and the $N_R$ have lepton number $-1$.

Some specific features of the framework in this paper are the following:
\begin{itemize}
\item $X$ and $Y$ are both
(in an appropriate weak basis) diagonal.
Therefore,
lepton mixing is induced solely
(in that weak basis)
by the charged-lepton mass matrix $M_\ell$ and by $M$. 
\item While $X$,
$Y$,
and $M_\ell$ arise from spontaneous symmetry breaking,
via Yukawa couplings to scalar fields
and via the vacuum expectation values (VEVs) of those fields,
$M$ is simply the matrix of the bare Majorana masses of the $N_R$.
We break one of the flavour symmetries \emph{softly}\/ at the
scale $m_{\rm soft}$ through the mass terms of the $N_R$,
a process which gives us enough freedom to enforce desired features
in the lepton mixing matrix.
The scale $m_{\rm soft}$ is \emph{naturally small}\/
because in the limit $M = 0$ a flavour symmetry is restored.
\item There must also be dimension-two terms which break softly
the flavour symmetry in the scalar potential.
Those terms are also assumed to be at a low scale of order $m_{\rm soft}$.
This renders \emph{vacuum alignment}\/
at the high (seesaw) scale $m_X$ natural,
apart from small corrections suppressed by $m_{\rm soft} / m_X$.
That vacuum alignment corresponds to the non-breaking by the vacuum
of one of the flavour symmetries.
\end{itemize}

This paper is organized as follows.
In section~\ref{model-tm1} we introduce a model for TM$_1$
and discuss its salient features.
Then,
by slight variations of the symmetries of the model and of the soft breaking,
we show in section~\ref{generalization}
how to realize other mixing schemes
like the ones in refs.~\cite{residual,lindner}.
Our conclusions are presented in section~\ref{concl}.
The precise determination of the flavour symmetry group
of the TM$_1$ model of section~\ref{model-tm1}
is relegated to appendix~A;
a discussion of some aspects of the scalar potential of that model
is undertaken 
in appendices~\ref{apb} and~\ref{apc}.

\section{A model for TM$_1$}
\label{model-tm1}
In this section we present a model for TM$_1$.
The leptonic multiplets of the model,
and also of all other models in this paper,
are the usual Standard-Model doublets $D_{jL}$ and
charged-lepton singlets $\ell_{jR}$,
together with the six right-handed neutrinos $\nu_{jR}$ and $N_{jR}$.
The scalar sector comprises \emph{four}\/ Higgs doublets,
$\phi_0$ and $\phi_j$,
and three complex singlets $S_j$.
In a concise notation,
for each type of field we subsume the three fields in column vectors:
\bs
\label{triplets}
\ba
&
D_L = \left( \begin{array}{c} D_{1L} \\ D_{2L} \\ D_{3L} \end{array} \right), 
\quad
\ell_R = \left( \begin{array}{c} \ell_{1R} \\ \ell_{2R} \\ \ell_{3R}
\end{array} \right), 
\quad 
\nu_R = \left( \begin{array}{c} \nu_{1R} \\ \nu_{2R} \\ \nu_{3R} 
\end{array} \right),
\quad
N_R = \left( \begin{array}{c} N_{1R} \\ N_{2R} \\ N_{3R} 
\end{array} \right),
& \hspace*{3mm}
\\*[2mm]
&
\phi = \left( \begin{array}{c} \phi_1 \\ \phi_2 \\ \phi_3 \end{array} \right),
\quad
S = \left( \begin{array}{c} S_1 \\ S_2 \\ S_3 \end{array} \right).
&
\ea
\es

For presenting the flavour symmetries of the model
it is expedient to define the unitary matrices 
\bs
\ba
E = \left( \begin{array}{ccc} 0 & 1 & 0 \\ 0 & 0 & 1 \\ 1 & 0 & 0
\end{array} \right),
& &
A = \left( \begin{array}{ccc}
1 & 0 & 0 \\ 0 & \omega & 0 \\ 0 & 0 & \omega^2
\end{array} \right),
\label{EA} \\
B = \left( \begin{array}{ccc}
1 & 0 & 0 \\ 0 & 0 & 1 \\ 0 & 1 & 0
\end{array} \right), 
& &
D = \left( \begin{array}{ccc}
1 & 0 & 0 \\ 0 & 0 & -1 \\ 0 & -1 & 0
\end{array} \right). 
\label{BBprime}
\ea
\es
With these matrices,
the symmetries of the TM$_1$ model,
acting on the flavour triplets in eqs.~(\ref{triplets}) and on $\phi_0$,
are formulated in table~\ref{tm1}.
\begin{table}[t]
\renewcommand{\arraystretch}{1.2}
\begin{center}
\begin{tabular}{c|ccccccc}
  & $D_L$ & $\ell_R$ & $\nu_R$ & $N_R$ & $S$ & $\phi$ & $\phi_0$ 
  \\ \hline
$\zzz$  & $E$ & $E$ & $E$ & $E$ & $E$ & $\bone$ & 1 \\
$\zzz'$ & $A$ & $A$ & $A$ & $A^\ast$ & $\bone$ & $A$ & $1$ \\
$\zz$   & $B$ & $B$ & $B$ & $D$ & $D$ & $B$ & $1$ \\
$\zzzz$ & $1$ & $1$ & $i$ & $1$ & $i$ & $1$ & $i$ 
\end{tabular}
\end{center}
\caption{Transformation properties of the multiplets
under the symmetries of the TM$_1$ model.}
\label{tm1}
\end{table}

Clearly,
the symmetry group of the model is $G = G' \times \mathbb{Z}_4$,
where $G'$ is the group generated by $\zzz$,
$\zzz'$,
and $\zz$.\footnote{In appendix~\ref{app} we prove that $G'$ is $\Delta(216)$.}
The Higgs doublet $\phi_0$ is invariant under $G'$ and can,
indeed,
act in our model like the Standard-Model Higgs doublet
which gives mass to the quarks.\footnote{The doublet $\phi_1$
is also invariant under $G'$ and constitutes another candidate
for the Standard-Model Higgs doublet.}

The symmetries in table~1 lead to the Yukawa Lagrangian 
\bs
\label{LY}
\ba
\mathcal{L}_\mathrm{Y} &=&
- y_0 \left( \phi_0^0,\, - \phi_0^+ \right) \sum_{j=1}^3 \bar \nu_{jR} D_{jL}
\label{LYphi0} \\ & &
- y_1 \left( \sum_{j=1}^3 \bar D_{jL} \ell_{jR} \right) \phi_1
\label{y1} \\ & &
- y_2 \left[
\left(
\bar D_{1L} \ell_{3R} + \bar D_{2L} \ell_{1R} + \bar D_{3L} \ell_{2R}
\right) \phi_2
\right. \no & & \left. \hspace*{5mm}
+ \left(
\bar D_{1L} \ell_{2R} + \bar D_{2L} \ell_{3R} + \bar D_{3L} \ell_{1R}
\right) \phi_3 \right]
\label{y2} \\ & &
- y_3 \sum_{j=1}^3 S_j \bar N_{jR} C \bar\nu_{jR}^T
+ \mbox{H.c.}
\label{LS}
\ea
\es
The symmetry $\zzzz$ is needed in order for the $\nu_{jR}$
to have Yukawa couplings to $\phi_0$ but not to the $\phi_j$.
That symmetry moreover impedes bare Majorana mass terms of the form
$\bar \nu_{jR} C \bar \nu_{kR}^T$.
The symmetry $\zz$ forbids extra terms
\bs
\label{forbid}
\ba
\label{forbid1}
& &
S_2 \bar N_{1R} C \bar \nu_{1R}^T
+ S_3 \bar N_{2R} C \bar \nu_{2R}^T
+ S_1 \bar N_{3R} C \bar \nu_{3R}^T,
\\
\label{forbid2}
& &
S_3 \bar N_{1R} C \bar \nu_{1R}^T
+ S_1 \bar N_{2R} C \bar \nu_{2R}^T
+ S_2 \bar N_{3R} C \bar \nu_{3R}^T
\ea
\es
in $\mathcal{L}_\mathrm{Y}$.

The mass terms of the charged leptons
\be
\mathcal{L}_{\ell\, \mathrm{mass}} = - \bar \ell_L M_\ell \ell_R + \mathrm{H.c.}
\ee
arise when the neutral components $\phi_j^0$ of the Higgs doublets $\phi_j$
acquire VEVs
$v_j \equiv \left\langle 0 \left| \phi_j^0 \right| 0 \right\rangle$.
One obtains the charged-lepton mass matrix
\be
M_\ell = y_1 v_1 \bone + y_2 \left( v_2 E^2 + v_3 E \right).
\ee
For the diagonalization of $M_\ell$ we use
the matrix $U_\omega$ of eq.~(\ref{Uomega}).
Since $U_\omega E U_\omega^\dagger = A^\ast$,
\be
U_\omega M_\ell U_\omega^\dagger = 
\mbox{diag} \left( x_e, x_\mu, x_\tau \right)
\ee
with 
\bs
\label{x}
\ba
x_e &=& y_1 v_1 + y_2 \left( v_2 + v_3 \right),
\\
x_\mu &=& y_1 v_1 + y_2 \left( \omega v_2 + \omega^2 v_3 \right),
\\
x_\tau &=& y_1 v_1 + y_2 \left( \omega^2 v_2
+ \omega v_3 \right).
\ea
\es
Clearly,
$v_2 \neq v_3$ is required
in order to have three different charged-lepton masses
$m_\alpha = | x_\alpha |$
($\alpha = e, \mu, \tau$).
That needs not pose a problem,
since the scalar potential
is sufficiently rich to enable $v_2 \neq v_3$ at its minimum,
as is demonstrated in appendix~\ref{apb}.

Recalling the definition of the matrices $X$ and $Y$ in eqs.~(\ref{zero}),
we find from the Yukawa Lagrangian the exceedingly simple forms
\bs
\ba
Y &=& y_0 v_0 \bone,
\\
X &=& y_3\, \mbox{diag} \left( s_1, s_2, s_3 \right),
\ea
\es
where
$v_0 \equiv  \left\langle 0 \left| \phi_0^0 \right| 0 \right\rangle$
and $s_j \equiv \left\langle 0 \left| S_j \right| 0 \right\rangle$.
In principle,
in a full model $\phi_0$ will be the Higgs doublet giving mass to the quarks.
Therefore,
$v_0$ will be of order the Fermi scale $m_{\rm Fermi} \sim$ 100 GeV.
Thus,
$m_Y \sim m_{\rm Fermi}$ or smaller, closer to the masses of the
charged leptons.

We may introduce bare neutrino Majorana mass terms for the $N_{jR}$ only:
\be
\mathcal{L}_{\mathrm{M}} =
- \frac{1}{2} \bar N_R M C \bar N_R^T 
+ \mathrm{H.c.}
\label{LM}
\ee
These bare Majorana mass terms have dimension three
and we allow them to \emph{softly}\/ break $\zzz$ and $\zzz'$
while \emph{preserving $\zz$}.\footnote{Softly breaking
the symmetries $\zzz$ and $\zzz'$
while leaving the symmetry $\zz$ intact
is an \emph{ad hoc}\/ assumption;
we make it solely because it leads to a viable and interesting model.}
(Note that the $N_{jR}$ transform trivially under $\zzzz$,
hence that symmetry cannot constrain the mass matrix $M$,
but it does forbid bare Majorana mass terms of the $\nu_{jR}$.)
Therefore,
\be
M = \left( \begin{array}{ccc}
a+2b & f & -f \\ f & a-b & d \\ - f & d & a-b
\end{array} \right),
\label{M}
\ee
with free mass parameters $a$,
$b$,
$d$,
and $f$.\footnote{We use the same notation for the mass parameters
as in ref.~\cite{varzielas}.}

As stressed in the introduction,
we assume the soft breaking of the symmetries
in both $\mathcal{L}_{\mathrm{M}}$ and the scalar potential to be \emph{small}\/
(relative to the Fermi scale).
The VEVs $s_j$ define the seesaw scale $m_X$ of $X$,
which is---just as in the standard seesaw mechanism---assumed to be
much higher the $m_{\rm Fermi}$.
Thus,
$s_j \gg m_{\rm soft}$ and it is legitimate to assume that
the $s_j$ are only very slightly perturbed
by the breaking of $\zzz$ at the scale $m_{\rm soft}$.
We may then assume $s_1 = s_2 = s_3 \equiv s$,
\textit{i.e.}~that the symmetry $\zzz$
remains unbroken at the seesaw scale.\footnote{In appendix~\ref{apc}
we discuss the scalar potential of the $S_j$.}
Then,
our seesaw formula~(\ref{mnu}) yields
\be
M_\nu = \frac{y_0^2}{y_3^2}\, \frac{v_0^2}{s^2}\, M. 
\ee
Therefore,
the unitary matrix $V$ which diagonalizes $M$,
\be
\label{V}
V^T\! M V = \hat M \quad {\rm with}\ \hat M\ {\rm diagonal},
\ee
also diagonalizes $M_\nu$.
Consequently,
the lepton mixing matrix $U$ is given by the product in eq.~(\ref{UV}).

Now,
\be
u = \frac{1}{\sqrt{2}} \left( \begin{array}{c} 0 \\ 1 \\ 1 
\end{array} \right)
\ee
is an eigenvector of $M$ with eigenvalue $a-b+d$. 
Therefore,
$u$ is a column vector of $V$.
Since~\cite{varzielas,tm1-mechanism} 
\be
U_\omega u = u_1,
\ee
we conclude that $u_1$ is a column vector of $U$.
This is precisely what one needs for TM$_1$.

\section{Generalization}
\label{generalization}
\subsection{Variations on the flavour symmetry}
In the TM$_1$ model an essential ingredient is the $\zz$ symmetry,
which forbids the Yukawa couplings in eqs.~(\ref{forbid})
and shapes the mass matrix $M$
in such a way that one can achieve the desired form of $U$.
We may change the symmetry $\zz$ and thus change the shape of $M$,
but we have to ensure that the new symmetry $\zz$
still forbids the terms in eqs.~(\ref{forbid}).
We firstly consider two types of symmetries
under which the Yukawa Lagrangian of eq.~(\ref{LY}) is invariant.
The first type of symmetries is
\be
\mbox{type (a):}
\quad
S_j \to e^{i\alpha_j} S_j,
\quad
N_{jR} \to e^{i\alpha_j} N_{jR},
\ee
where the phases $\alpha_j$ are arbitrary
and all the other fields transform trivially.
The second type of symmetries is 
\be
\mbox{type (b):} \quad \left\{ 
\begin{array}{ccc}
S_1 \to e^{i\alpha_1} S_1, &
S_2 \to e^{i\alpha_3} S_3, &
S_3 \to e^{i\alpha_2} S_2,
\\
N_{1R} \to e^{i\alpha_1} N_{1R}, &
N_{2R} \to e^{i\alpha_3} N_{3R}, &
N_{3R} \to e^{i\alpha_2} N_{2R},
\\
\nu_{2 R} \leftrightarrow \nu_{3 R}, \\*[1mm]
D_{2 L} \leftrightarrow D_{3 L}, \\*[1mm]
\ell_{2R} \leftrightarrow \ell_{3R}, \\*[1mm]
\phi_2 \leftrightarrow \phi_3,
\end{array} \right.
\ee
where once again the phases $\alpha_j$ are arbitrary.
Secondly,
we require that
these transformations eliminate the terms of eqs.~(\ref{forbid});
this happens provided the phases $\alpha_j$ are not all equal.
Finally,
we require that the above symmetries are of the $\zz$ type,
so that they may constitute an invariance of $M$;
this happens if the phases $\alpha_j$ are either 0 or $\pi$
for symmetries of type (a),
and if $\exp{\left( i \alpha_1 \right)} = \pm 1$,
$\exp{\left[ i \left( \alpha_2 + \alpha_3 \right) \right]} = 1$
for symmetries of type (b).
For instance,
the $\zz$ symmetry of the TM$_1$ model is
\be
\label{b}
{\rm type\ (b)\ with}\ e^{i\alpha_1} = +1,\
e^{i\alpha_2} = e^{i\alpha_3} = -1.
\ee

An alternative $\zz$ symmetry that we might impose would be
\be
\label{a}
{\rm type\ (a)\ with}\ e^{i\alpha_1} = +1,\
e^{i\alpha_2} = e^{i\alpha_3} = -1.
\ee
This renders the matrix $M$ block-diagonal,
with the 
Cartesian 
basis vector $e_1$ being one of its eigenvectors.
Then,
$e_1$ is a column in $V$ and,
according to eq.~(\ref{UV}),
trimaximal mixing,
{\it i.e.}~TM$_2$,
ensues.
We have thus constructed a model for TM$_2$.

Clearly,
one can also envisage the imposition of two $\zz$ symmetries
instead of only one.
For instance,
imposing both the symmetries of eqs.~(\ref{b}) and~(\ref{a})
leads to simultaneous TM$_1$ and TM$_2$.
We thus have a model for TBM,
which however is now phenomenologically ruled out.

\subsection{Predicting the reactor mixing angle}
We consider in this subsection the following generalization of eq.~(\ref{b}):
\be
\label{z2}
{\rm type\ (b)\ with}\quad e^{i\alpha_1} = +1,\quad
e^{i\alpha_2} = e^{- i\alpha_3} = e^{i \alpha} \neq \pm 1.
\ee
With this choice one obtains
\be
M = \left( \begin{array}{ccc}
M_{11} & M_{12} & M_{12}  e^{i\alpha} \\
M_{12} & M_{22} & M_{23} \\
M_{12} e^{i\alpha} & M_{23} & M_{22} e^{2i\alpha}
\end{array} \right).
\label{newM}
\ee
It is easy to find a column vector $u$ of the matrix $V$ which diagonalizes $M$.
According to eq.~(\ref{V}),
such a vector must have the property $M u \propto u^*$.
So it is given in the case of the $M$ of eq.~(\ref{newM}) by 
\be
u = \frac{1}{\sqrt{2}} \left( 
\begin{array}{c} 0 \\ 1 \\ -e^{-i\alpha} 
\end{array} \right),
\ee
since $M u = \left( M_{22} - M_{23} e^{-i\alpha} \right) u^*$.
Therefore,
\be
\label{Uu}
U_\omega u = \frac{1}{\sqrt{6}} \left(
\begin{array}{c}
1 - e^{-i\alpha} \\
\omega - \omega^2 e^{-i\alpha} \\
\omega^2 - \omega e^{-i\alpha} 
\end{array} \right)
\ee
will be one of the columns of the mixing matrix $U$.
This is a generalization of the TM$_1$ model 
of the previous section.
In this generalization,
$\alpha$ is some well-defined phase.

Which column of $U$ is $U_\omega u$?
This depends on the value of $\alpha$.
If $e^{i\alpha}$ were 1,\footnote{This choice 
does not eliminate the terms of eqs.~(\ref{forbid}),
so in this case one would have to impose some further
symmetry in order to get rid of those terms.}
then $U_\omega u$ would have a zero element;
this indicates that,
in the closest approximation to the \emph{phenomenological}\/
lepton mixing matrix,
$U_\omega u$ should be the third column of $U$,
yielding $s_{13}^2 = 0$ and $s_{23}^2 = 1/2$.
This is of course now ruled out,
because we know that $s_{13} \neq 0$.
For $e^{i\alpha} = -1$
one should choose $U_\omega u$ to be the first column of $U$,
reproducing TM$_1$ as we have seen in the previous section of this paper.

\begin{figure}[t]
\begin{center}
\epsfig{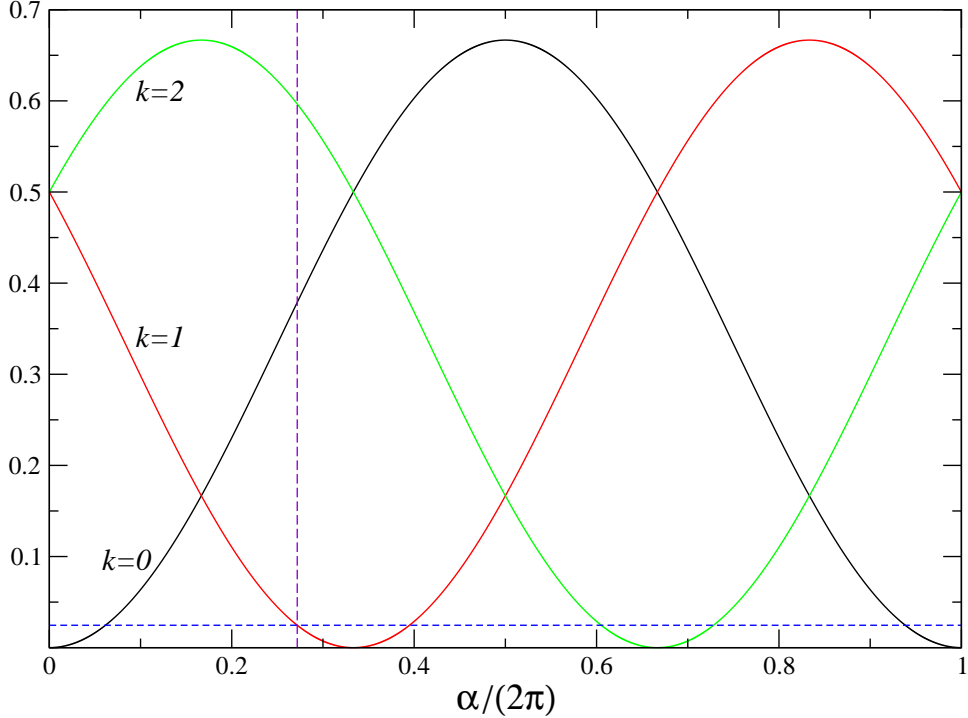}
\end{center}
\caption{The functions $f_k (\alpha)$ of eq.~(\ref{fk}).
The dashed horizontal line indicates $s_{13}^2 = 0.0246$,
the best-fit value of Forero \textit{et al.}~\cite{schwetz}.
\label{fig}
}
\end{figure}

Let us in the following assume that $U_\omega u$
is the {\em third}\/ column of $U$.
Still,
we can permute the order of the charged-lepton masses in eq.~(\ref{x}).
This means that $s_{13}^2$ could be the squared modulus
of any of the elements of $U_\omega u$ in eq.~(\ref{Uu}):
\be
\label{fk}
s_{13}^2 = \frac{1}{6} \left| 1- \omega^k e^{-i\alpha} \right|^2
\equiv f_k (\alpha),
\ee
with $k$ being either 0,
1,
or 2.
The squared moduli of the other two elements of $U_\omega u$
must then be identified with $c_{13}^2 s_{23}^2$ and $c_{13}^2 c_{23}^2$. 
We have plotted the functions $f_k$ in fig.~\ref{fig}.
In that figure we also displayed a dashed horizontal line which indicates
a phenomenologically realistic value of $s_{13}^2$.
That horizontal line intersects each curve $f_k$
for two distinct values of $\alpha$.
As an example,
the vertical line in the figure shows
the intersection point $\alpha = 0.2718 \left( 2 \pi \right)$ of $f_1$;
in this example we have $s_{13}^2 = f_1 (\alpha)$,
and then $c_{13}^2 s_{23}^2$ would be either $f_0 (\alpha)$ or $f_2 (\alpha)$.

\begin{figure}[t]
\begin{center}
\epsfig{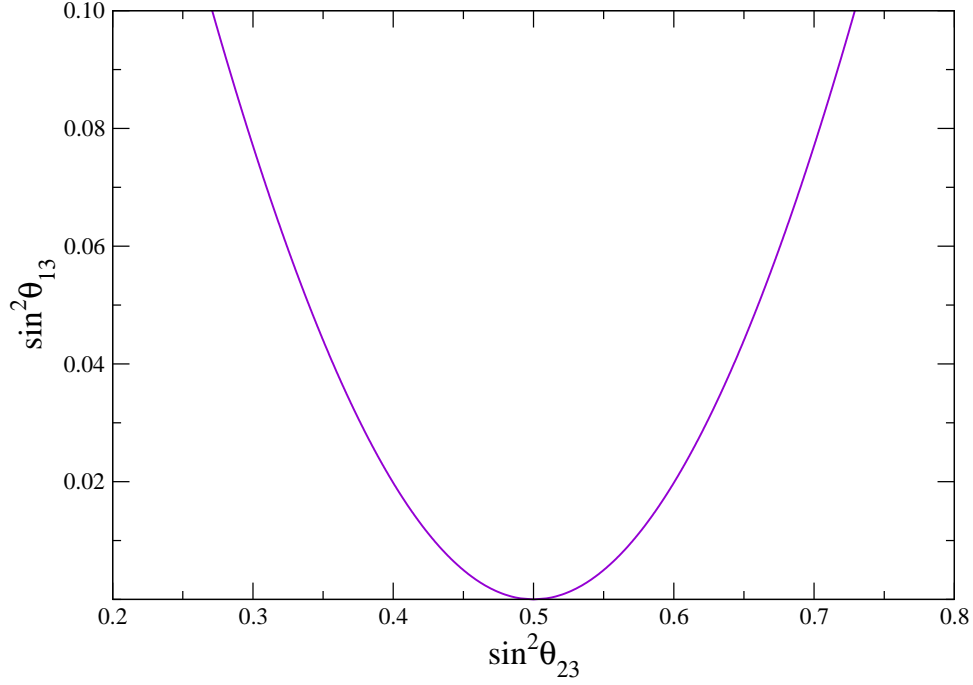}
\end{center}
\caption{The relation between $s_{13}^2$ and $s_{23}^2$,
a graphical rendering of eq.~(\ref{s23}).
\label{parabo}}
\end{figure}

We can read off two facts from fig.~\ref{fig}.
Firstly,
for every value of $s_{13}^2$ there are two possible values of $s_{23}^2$.
Secondly,
all the intersection points lead to an identical relation
between $s_{13}^2$ and $s_{23}^2$,
{\it i.e.}~the two values of $s_{23}^2$ are always the same
no matter which $k$-curve one has chosen.
Therefore,
for simplicity we can take $k=0$ in eq.~(\ref{fk}).
Analytically,
one then finds the relation
\be
\left( 1 - s_{13}^2 \right)^2 \left( 2 s_{23}^2 - 1 \right)^2
= 2 s_{13}^2 - 3 s_{13}^4.
\ee
Solving this equation for $s_{23}^2$ yields the two solutions 
\be
\label{s23}
s_{23}^2 = \frac{1}{2} \left( 1 \pm 
\frac{\sqrt{2 s_{13}^2 - 3 s_{13}^4}}{c_{13}^2} \right).
\ee
This is plotted in fig.~\ref{parabo}.
For definiteness in that figure we have allowed $s_{13}^2$
to vary between zero and $0.1$;
in this limited range,
the curve is almost a parabola~\cite{lindner}.

Of course,
in any definite model within our framework
we must choose a well-defined value for $\alpha$,
and thereby choose one point in the pseudo-parabola of fig.~2.
We may,
in particular,
require that $\alpha$ is such that
the model has a {\em finite}\/ flavour symmetry group;
a necessary condition for this is that $e^{i\alpha}$ should be a root of unity. 
In this case $\alpha / (2\pi)$ must be a rational number
which approximates well
one of the intersection points for the phenomenological value of $s_{13}^2$.
In particular,
the values of $\alpha$ in table~2
reproduce the phenomenological data quite well,
as was first found in ref.~\cite{lindner}.
\begin{table}[ht]
\renewcommand{\arraystretch}{1.2}
\begin{center}
\begin{tabular}{|c|c|c|}
\hline
$\alpha \left/ \left( 2 \pi \right) \right. $ & $s_{13}^2$ & $s_{23}^2$ \\ \hline
2/5, 3/5 & 0.028818 & 0.379101 or 0.620899 \\
1/16, 15/16 & 0.025373 & 0.386653 or 0.613347 \\
1/18, 5/18, 7/18, 11/18, 13/18, 17/18 & 0.020102 & 0.399242
or 0.600758 \\ \hline
\end{tabular}
\end{center}
\caption{Some rational values
of $\alpha \left/ \left( 2 \pi \right) \right. $ for which $s_{13}^2$,
computed by using eq.~(\ref{fk}),
turns out to agree with the phenomenological data.
The two corresponding values of $s_{23}^2$ follow
from each value of $s_{13}^2$ according to eq.~(\ref{s23}).}
\label{tm2}
\end{table}

Furthermore,
one may additionally require trimaximal mixing by using the additional,
and independent,
symmetry of eq.~(\ref{a}).
Then the solar mixing angle is obtained from
$s_{12}^2 = 1 \left/ \left( 3 c_{13}^2 \right) \right. $.
The requirement of TM$_2$ determines the $CP$-violating phase $\delta$
as~\cite{we} 
\be
\cos{\delta} = \frac{\left( 1 - 2 s_{13}^2 \right)
\left( 1 - 2 s_{23}^2 \right)}
{2 s_{13} s_{23} c_{23} \sqrt{2 - 3 s_{13}^2}}\, .
\ee
However, 
taking into account eq.~(\ref{s23}), we simply find
\be
\cos{\delta} = \mp 1,
\ee
where the upper (lower) sign corresponds to the upper (lower) sign 
in eq.~(\ref{s23}). 
Thus, the $\zz$ symmetry~(\ref{z2}) together with TM$_2$ leads to 
$\delta = 0$ or $\pi$, as was noticed for the viable cases of $U$
studied in ref.~\cite{lindner}. 

Since in the case discussed here we have determined two columns of $U$,
the whole mixing matrix $U$
becomes,
apart from possible permutations of the rows,
determined as a function of $\alpha$:
\be\label{Ua}
\renewcommand{\arraystretch}{1.5}
U = \frac{1}{\sqrt{6}} \left( \begin{array}{ccccc}
1 + e^{i\alpha} & &
\sqrt{2} & &
1 - e^{-i\alpha}
\\
\omega^2 + \omega e^{i\alpha} & &
\sqrt{2} & &
\omega - \omega^2 e^{-i\alpha}
\\
\omega + \omega^2 e^{i\alpha} & &
\sqrt{2} & &
\omega^2 - \omega e^{-i\alpha}
\end{array} \right).
\ee
One may use this mixing matrix to check again
that the $CP$-violating phase $\delta$
turns out to be trivial.
However, the Majorana phases are non-trivial
functions of $\alpha$, as can be read off 
from the following forms of
the $k$-th entries in the first and third column:
\be
U_{k1} = \sqrt{\frac{2}{3}} \cos \left( \frac{\alpha}{2} - 
\frac{2\pi k}{6} \right) e^{i \alpha/2} (-1)^k, 
\quad
U_{k3} = \sqrt{\frac{2}{3}} \sin \left( \frac{\alpha}{2} - 
\frac{2\pi k}{6} \right) i\, e^{-i \alpha/2} (-1)^k.
\ee

\section{Conclusions}
\label{concl}

In this paper we have introduced
a class of renormalizable models
based on the double seesaw mechanism
and on the soft breaking of flavour symmetries.
In order to implement the double seesaw mechanism,
the models possess
three right-handed singlet fields $N_{jR}$,
in addition to the $\nu_{jR}$ needed for the
usual seesaw mechanism.
Moreover, the models have
an enlarged scalar sector compared to the Standard Model,
namely four Higgs doublets and three complex scalar singlets.
We stress that this is a rather minor field content
in comparison to usual scenarios in model building
with flavour symmetries, especially when those scenarios are supersymmetric.

Our class of models has both spontaneous and soft symmetry breaking.
Spontaneous symmetry breaking occurs at two scales: at the scale
$m_X$ through the VEVs of the scalar gauge singlets
and at the Fermi scale through the VEVs of the Higgs doublets. 
Soft flavour symmetry breaking
happens in the mass terms of the $N_{jR}$ at a scale $m_\mathrm{soft}$.
If we assume that $m_\mathrm{soft}$ is much smaller than the Fermi scale,
then the spontaneous symmetry breaking
proceeds nearly unperturbed by the soft symmetry breaking.
Due to the double seesaw mechanism, the mass scale $m_\nu$
of the light neutrinos is determined by
\be\label{scales}
m_\nu \sim \frac{m_Y^2}{m_X^2}\,m_\mathrm{soft},
\ee
where $m_Y$ is at most of the order of the Fermi scale
but might also be much smaller, since
the masses of the charged leptons are considerably smaller than the Fermi scale.
With $m_\nu \sim$ 0.1 eV,
eq.~(\ref{scales}) permits an estimate of $m_\mathrm{soft}/m_X^2$,
but no independent determination of the soft-breaking
scale and of the seesaw scale $m_X$.

Our flavour symmetries are arranged
in such a way that, in an appropriate weak basis,
the contribution to the lepton mixing matrix $U$
from the charged-lepton sector is given by $U_\omega$ of eq.~(\ref{Uomega}),
whereas the neutrino sector contributes a matrix $V$
constrained by either one or two $\zz$ symmetries~\cite{residual}.
The matrix $V$ is exclusively determined
by the Majorana mass matrix $M$ of the $N_{jR}$.
Our class of models allows one
to impose $\zz$ symmetries
which lead to 
virtually 
any of the forms of $U$ that have been
proposed in the literature~\cite{residual,lindner}.
This arbitrariness 
may be viewed as a weak point of our models,
which on the other hand have the advantage of being renormalizable
and natural in a technical sense.

We have explicitly discussed a model for TM$_1$. 
Then,
by variation of the $\zz$ symmetry of that model
but keeping all its other flavour symmetries intact,
we have shown that one can also achieve either TM$_2$
or a determination of the third column of $U$.
The latter is enforced through a $\zz$ symmetry
depending on an angle $\alpha$.
By varying $\alpha$,
different values of $s_{13}^2$ are produced.
In this way,
we have explicitly reproduced the values of $s_{13}^2$ found
in ref.~\cite{lindner}. 

Finally, 
combination of TM$_2$ and of the $\zz$ symmetry depending on $\alpha$
results in the one-parameter mixing matrix of eq.~(\ref{Ua}),
whose predictions we have discussed.\footnote{Recently~\cite{feruglio},
other one-parameter mixing matrices
have been obtained from residual flavour and $CP$ symmetries.} 

\begin{appendix}

\setcounter{equation}{0}
\renewcommand{\theequation}{A\arabic{equation}}

\section{The flavour symmetry group of the TM$_1$ model}
\label{app}

We firstly recall the definition
of the group series $\Delta(6n^2)$,
where $n$ is an integer.
The easiest way to understand these groups is to conceive them as
the subgroups of $SU(3)$ generated by
\be
r = E, \quad s = -B, \quad t = 
T \equiv \mbox{diag} \left( 1,\, \eta,\, \eta^\ast \right)
\quad \mbox{with} \quad
\eta \equiv \exp{\left( 2 \pi i / n \right)}.
\ee
The definition of $\Delta(6n^2)$ in ref.~\cite{luhn}
uses a different set of generators $a$, $b$, $c$,
and $d$ related to ours through 
$r = a$, $b = r^{-1}sr$, $c = rtr^{-1}$, and $d = t$.

Every group $\Delta(6n^2)$ has 
$2(n-1)$ three-dimensional inequivalent irreducible representations
(irreps) $\three^{(k \pm)}$ ($k = 1, \ldots, n-1$),
which are given in an appropriate basis by~\cite{luhn}
\be
\label{irreps}
\three^{(k \pm)}: \quad
r \to E, \quad s \to \mp B, \quad
t \to \mbox{diag} \left( 1,\, \eta^k,\, \eta^{-k} \right).
\ee

The flavour symmetry group of our TM$_1$ model is $G = G' \times \zzzz$.
The group $G'$ is the one generated by three transformations $g_{1,2,3}$.
We know the following three representations of those transformations
(in the first five columns of table~1):
\bs
\label{irrepsss}
\ba
D_L, \ell_R, \nu_R: & & g_1 \to E,\ g_2 \to A,\ g_3 \to B,
\\
N_R: & & g_1 \to E,\ g_2 \to A^\ast,\ g_3 \to D,
\\ 
S: & & g_1 \to E,\ g_2 \to \bone,\ g_3 \to D.
\ea
\es
The Higgs doublet $\phi_0$ is invariant under $G'$.
We leave the three Higgs doublets $\phi_j$ for later consideration.

Let us define a particular transformation $g \in G'$
through $g \equiv g_1 (g_1 g_3)^2 g_1^{-1}$.
One readily ascertains that
\bs
\ba
E \left( E B \right)^2 E^{-1} &=& \bone,
\label{one}
\\
E \left( E D \right)^2 E^{-1} &=& W \equiv {\rm diag} \left( 1, -1, -1 \right).
\label{W}
\ea
\es
The unit matrix therefore represents $g$ in the representation of $D_L$,
$\ell_R$,
and $\nu_R$,
while $W$ represents $g$ in the representation of $N_R$
and in the representation of $S$.
Let us now define two further transformations $g'_{2,3} \in G'$
through $g'_{2,3} \equiv g\, g_{2,3}$.
Since
\bs
\ba
W D &=& B, \\
W A^\ast &=& {\rm diag} \left( 1,\, e^{2 \pi i / 6},\, e^{- 2 \pi i / 6} \right),
\ea
\es
one concludes that the representations in eqs.~(\ref{irrepsss})
might as well be given through
\bs
\label{irrepssss}
\ba
D_L, \ell_R, \nu_R: & &
g_1 \to E,\
g'_2 \to {\rm diag} \left( 1,\, e^{2 \pi i / 3},\, e^{- 2 \pi i / 3} \right),\
g'_3 \to B,
\\
N_R: & &
g_1 \to E,\
g'_2 \to {\rm diag} \left( 1,\, e^{2 \pi i / 6},\, e^{- 2 \pi i / 6} \right),\
g'_3 \to B,
\\ 
S: & & g_1 \to E,\
g'_2 \to {\rm diag} \left( 1,\, -1,\, -1 \right),\
g'_3 \to B.
\ea
\es
We conclude that $G' = \Delta (216)$ with
\bs
\ba
D_L, \ell_R, \nu_R: & & {\mathbf 3}^{(2-)},
\\
N_R: & & {\mathbf 3}^{(1-)},
\\
S: & & {\mathbf 3}^{(3-)}.
\ea
\es

Actually,
the argument for $\Delta(216)$,
as developed above,
boils down to the following.
The symmetries $\zzz$ and $\zz$
together generate a group $S_4$;
this can be seen by considering the action of $\zzz$ and $\zz$
on $N_R$ and on $S$.
However,
there is also $\zzz'$.
This suggests that
\be
G' \cong \left( \zzz' \times \zzz' \right) \rtimes S_4
\cong \left( \z_6 \times \z_6 \right) \rtimes S_3
\cong \Delta \left( 6 \times 6^2 \right)
= \Delta(216),
\ee
where we have used $S_4 \cong \left (\zz \times \zz \right) \rtimes S_3$
and $\z_6 \cong \zzz' \times \zz$.

We lastly investigate the representation of $\phi$.
This multiplet obviously transforms under $G'$ as ${\bf 1} \oplus {\bf 2}$,
where ${\bf 1}$ is invariant under $G'$
and the two-dimensional irrep is given by
\be
{\bf 2}: \quad
g_1 \to \left( \begin{array}{cc} 1 & 0 \\ 0 & 1 \end{array} \right),
\quad
g_2 \to \left( \begin{array}{cc} \omega & 0 \\ 0 & \omega^2 \end{array} \right),
\quad
g_3 \to \left( \begin{array}{cc} 0 & 1 \\ 1 & 0 \end{array} \right).
\ee
Therefore $g$ is represented in the ${\bf 2}$ by the $2 \times 2$ unit matrix
and
\be
{\bf 2}: \quad
g_1 \to \left( \begin{array}{cc} 1 & 0 \\ 0 & 1 \end{array} \right),
\quad
g'_2 \to
 \left( \begin{array}{cc} \omega & 0 \\ 0 & \omega^2 \end{array} \right),
\quad
g'_3 \to
 \left( \begin{array}{cc} 0 & 1 \\ 1 & 0 \end{array} \right).
\ee
This is one of the four two-dimensional universal
irreps\footnote{By ``universal'' we mean that
they do not depend on the precise value of $n$,
provided $n$ is divisible by 3.}
of $\Delta(6n^2)$ which exist whenever $n$ is a multiple of 3~\cite{luhn,GL11},
as is the case for $\Delta (216)$.\footnote{The present model
shares some similarities with the model of ref.~\cite{FGLL},
which is also based on the flavour group $\Delta (216)$.}

\setcounter{equation}{0}
\renewcommand{\theequation}{B\arabic{equation}}

\section{The $\phi_j$ potential}
\label{apb}

The symmetries $\zzz'$ and $\zz$ act on the Higgs doublets $\phi_j$
($j = 1, 2, 3$)
as if they constituted a (reducible) triplet of a group $S_3$.
Therefore,
the potential for those three doublets alone is
\ba
V_\phi &=&
\mu_1\, \phi_1^\dagger \phi_1
+ \mu_2 \left( \phi_2^\dagger \phi_2 + \phi_3^\dagger \phi_3 \right)
\no & &
+ \lambda_1 \left( \phi_1^\dagger \phi_1 \right)^2
+ \lambda_2 \left[ \left( \phi_2^\dagger \phi_2 \right)^2
+ \left( \phi_3^\dagger \phi_3 \right)^2 \right]
\no & &
+ \lambda_4\, \phi_1^\dagger \phi_1
\left( \phi_2^\dagger \phi_2 + \phi_3^\dagger \phi_3 \right)
+ \lambda_5\, \phi_2^\dagger \phi_2\, \phi_3^\dagger \phi_3
\no & &
+ \lambda_4^\prime \left( \phi_1^\dagger \phi_2\, \phi_2^\dagger \phi_1
+ \phi_1^\dagger \phi_3\, \phi_3^\dagger \phi_1 \right)
+ \lambda_5^\prime\, \phi_2^\dagger \phi_3\, \phi_3^\dagger \phi_2
\no & &
+ \left[
\lambda_6\, \phi_1^\dagger \phi_2\, \phi_1^\dagger \phi_3
+ \lambda_7 \left( \phi_1^\dagger \phi_2\, \phi_3^\dagger \phi_2
+ \phi_1^\dagger \phi_3\, \phi_2^\dagger \phi_3 \right)
+ {\rm H.c.} \right].
\ea
Let
$\tilde \lambda_4 \equiv \lambda_4 + \lambda_4^\prime$
and
$\tilde \lambda_5 \equiv \lambda_5 + \lambda_5^\prime$.
Then,
the VEV of the potential is
\ba
V_0 \equiv \left\langle 0 \left| V_\phi \right| 0 \right\rangle &=&
\mu_1 \left| v_1 \right|^2
+ \mu_2 \left( \left| v_2 \right|^2 + \left| v_3 \right|^2 \right)
+ \lambda_1 \left| v_1 \right|^4
+ \lambda_2 \left( \left| v_2 \right|^4 + \left| v_3 \right|^4 \right)
\no & &
+ \tilde \lambda_4 \left| v_1 \right|^2
\left( \left| v_2 \right|^2 + \left| v_3 \right|^2 \right)
+ \tilde \lambda_5 \left| v_2 v_3 \right|^2
\no & &
+ 2\, \Re \left[ \lambda_6 {v_1^\ast}^2 v_2 v_3
+ \lambda_7 v_1^\ast \left( v_2^2 v_3^\ast + v_3^2 v_2^\ast \right) \right].
\ea
It is hard to proceed analytically in the general case.
We shall for the sake of simplification assume $\lambda_6 = 0$,
even though there is no symmetry that supports that assumption.
When $\lambda_6 = 0$ the two relative phases among $v_1$,
$v_2$,
and $v_3$ adjust so that $\lambda_7 v_1^\ast v_2^2 v_3^\ast$
and $\lambda_7 v_1^\ast v_3^2 v_2^\ast$ are both real and negative.
One may then write
\be
\tilde V_0 =
\tilde \mu_2 \left( \left| v_2 \right|^2 + \left| v_3 \right|^2 \right)
+ \lambda_2 \left( \left| v_2 \right|^4 + \left| v_3 \right|^4 \right)
+ \tilde \lambda_5 \left| v_2 v_3 \right|^2
- \left| \tilde \lambda_7 \right|
\left( \left| v_2^2 v_3 \right| + \left| v_3^2 v_2 \right| \right),
\ee
where $\tilde V_0 \equiv V_0 - \mu_1 \left| v_1 \right|^2
- \lambda_1 \left| v_1 \right|^4$,
$\tilde \mu_2 \equiv \mu_2 + \lambda_4 \left| v_1 \right|^2$,
and $\tilde \lambda_7 \equiv 2 \lambda_7 v_1$.
The equations for vacuum stability are
\bs
\label{stabil}
\ba
\frac{\partial \tilde V_0}{\partial \left| v_2 \right|} = 0 &=&
2 \tilde \mu_2 \left| v_2 \right|
+ 4 \lambda_2 \left| v_2 \right|^3
+ 2 \tilde \lambda_5 \left| v_2 v_3^2 \right|
- 2 \left| \tilde \lambda_7 v_2 v_3 \right|
- \left| \tilde \lambda_7 v_3^2 \right|,
\label{v2} \\
\frac{\partial \tilde V_0}{\partial \left| v_3 \right|} = 0 &=&
2 \tilde \mu_2 \left| v_3 \right|
+ 4 \lambda_2 \left| v_3 \right|^3
+ 2 \tilde \lambda_5 \left| v_2^2 v_3 \right|
- 2 \left| \tilde \lambda_7 v_2 v_3 \right|
- \left| \tilde \lambda_7 v_2^2 \right|.
\label{v3}
\ea
\es
Subtracting the two eqs.~(\ref{stabil}) from each other,
we find that a solution with $\left| v_2 \right| \neq \left| v_3 \right|$
may exist provided
\bs
\label{fin}
\ba
2 \tilde \mu_2 &=&
- 4 \lambda_2 \left( \left| v_2 \right|^2 + \left| v_2 v_3 \right|
+ \left| v_3 \right|^2 \right)
+ 2 \tilde \lambda_5 \left| v_2 v_3 \right|
- \left| \tilde \lambda_7 \right| \left( \left| v_2 \right|
+ \left| v_3 \right| \right),
\\
0 &=& \left( 2 \tilde \lambda_5 - 4 \lambda_2 \right)
\left( \left| v_2^2 v_3 \right| + \left| v_2 v_3^2 \right| \right)
- \left| \tilde \lambda_7 \right| \left( \left| v_2 \right|^2
+ \left| v_3 \right|^2 + 3 \left| v_2 v_3 \right| \right).
\label{7}
\ea
\es
A solution to eqs.~(\ref{fin})
with $\left| v_2 \right|$ and $\left| v_3 \right|$ both positive
should exist for appropriate values of the parameters.
Notice the crucial role played 
by $\lambda_7$
in the existence of that solution---if $\lambda_7$ vanished
then $\tilde \lambda_5$ would have to be equal to $2 \lambda_2$
in order for eq.~(\ref{7}) to be satisfied;
but
$\tilde \lambda_5 = 2 \lambda_2$ is not stable under renormalization
because it is not supported by any extra symmetry of the potential. 

The stability point of $\tilde V_0$ with $v_2 \neq v_3$
will actually be a local minimum provided
the matrix of the second derivatives of $\tilde V_0$
relative to $\left| v_2 \right|$ and $\left| v_3 \right|$,
computed under the conditions of eqs.~(B5),
is positive definite. This means, apart from requiring positivity of
the determinant of that matrix,
we have to ensure that
\bs
\ba
4 \lambda_2 \left( 2 \left| v_2 \right|^2 - \left| v_2 v_3 \right|
- \left| v_3 \right|^2 \right)
+ 2 \tilde \lambda_5 \left( \left| v_3 \right|^2 + \left| v_2 v_3 \right| \right)
- \left| \tilde \lambda_7 \right| \left( \left| v_2 \right|
+ 3 \left| v_3 \right| \right) &>& 0,
\\
4 \lambda_2 \left( 2 \left| v_3 \right|^2 - \left| v_2 v_3 \right|
- \left| v_2 \right|^2 \right)
+ 2 \tilde \lambda_5 \left( \left| v_2 \right|^2 + \left| v_2 v_3 \right| \right)
- \left| \tilde \lambda_7 \right| \left( \left| v_3 \right|
+ 3 \left| v_2 \right| \right) &>& 0.
%\\
%4 \tilde \lambda_5 \left| v_2 v_3 \right| - 2 \left| \tilde \lambda_7 \right|
%\left( \left| v_2 \right| + \left| v_3 \right| \right) &>& 0.
\ea
\es
We have moreover to ensure that this local minimum
attains a lower value for $\tilde V_0$
than the solution to eqs.~(\ref{stabil})
with $\left| v_2 \right| = \left| v_3 \right|$.
In a more thorough study,
we would also have to look for possible minima of $\tilde V_0$
which break the electric-charge invariance.

\setcounter{equation}{0}
\renewcommand{\theequation}{C\arabic{equation}}

\section{The $S_j$ potential}
\label{apc}

The potential for the complex gauge singlets $S_j$
($j = 1, 2, 3$)
must be invariant under the symmetries $\zzz$,
$\zz$,
and $\zzzz$,
\textit{i.e.}\ under $S_1 \to S_2 \to S_3 \to S_1$,
under $S_2 \leftrightarrow - S_3$,
and under $S_j \to i S_j,\, \forall j$.
Therefore,
\ba
V_S &=&
\bar \mu \sum_{j=1}^3 \left| S_j \right|^2
+ \bar \lambda_1 \left( \sum_{j=1}^3 \left| S_j \right|^2 \right)^2
+ \bar \lambda_2 \left( \left| S_1 S_2 \right|^2 + \left| S_1 S_3 \right|^2
+ \left| S_2 S_3 \right|^2 \right)
\no & &
+ \left\{
\bar \lambda_3 \left( \sum_{j=1}^3 S_j^2 \right)^2
+ \bar \lambda_4 \left[
\left( S_1 S_2 \right)^2 + \left( S_1 S_3 \right)^2 + \left( S_2 S_3 \right)^2
\right]
{\rm + H.c.} \right\}
\no & &
+ \bar \lambda_5 \left[
\left( S_1^\ast S_2 \right)^2 + \left( S_2^\ast S_3 \right)^2
+ \left( S_3^\ast S_1 \right)^2
+ {\rm H.c.} \right],
\label{spot}
\ea
with complex $\bar \lambda_3$ and $\bar \lambda_4$ but real $\bar \lambda_5$.
The equations for vacuum stability are
\bs
\label{vevstab}
\ba
0 &=& \bar \mu s_1^\ast + 2 \bar \lambda_1 \left| s_1 \right|^2 s_1^\ast
+ \left( 2 \bar \lambda_1 + \bar \lambda_2 \right)
\left( \left| s_2 \right|^2 + \left| s_3 \right|^2 \right) s_1^\ast
\no & &
+ 4 \bar \lambda_3 s_1^3
+ \left( 4 \bar \lambda_3 + 2 \bar \lambda_4 \right)
\left( s_2^2 + s_3^2 \right) s_1
+ 2 \bar \lambda_5 \left( {s_2^\ast}^2 + {s_3^\ast}^2 \right) s_1,
\\
0 &=& \bar \mu s_2^\ast + 2 \bar \lambda_1 \left| s_2 \right|^2 s_2^\ast
+ \left( 2 \bar \lambda_1 + \bar \lambda_2 \right)
\left( \left| s_1 \right|^2 + \left| s_3 \right|^2 \right) s_2^\ast
\no & &
+ 4 \bar \lambda_3 s_2^3
+ \left( 4 \bar \lambda_3 + 2 \bar \lambda_4 \right)
\left( s_1^2 + s_3^2 \right) s_2
+ 2 \bar \lambda_5 \left( {s_1^\ast}^2 + {s_3^\ast}^2 \right) s_2,
\\
0 &=& \bar \mu s_3^\ast + 2 \bar \lambda_1 \left| s_3 \right|^2 s_3^\ast
+ \left( 2 \bar \lambda_1 + \bar \lambda_2 \right)
\left( \left| s_1 \right|^2 + \left| s_2 \right|^2 \right) s_3^\ast
\no & &
+ 4 \bar \lambda_3 s_3^3
+ \left( 4 \bar \lambda_3 + 2 \bar \lambda_4 \right)
\left( s_1^2 + s_2^2 \right) s_3
+ 2 \bar \lambda_5 \left( {s_1^\ast}^2 + {s_2^\ast}^2 \right) s_3.
\ea
\es

The potential $V_S$ has several accidental symmetries,
like for instance under $S_1 \to - S_1$ and under $S_1 \leftrightarrow S_2$.
Correspondingly,
solutions to eqs.~(\ref{vevstab}) may exist with varying features,
like $s_1 = 0$,
$s_1 = s_2 \neq 0$,
or $s_1 = s_2 = 0$.
In principle,
a solution to eqs.~(\ref{vevstab}) with the three $s_j$
all nonzero and distinct may also exist.

In this paper we \emph{assume}\/ that the parameters of the potential
are such that the solution to eqs.~(\ref{vevstab})
featuring $s_1 = s_2 = s_3 \equiv s$,
with
\be
0 = \bar \mu s^\ast
+ \left( 6 \bar \lambda_1 + 2 \bar \lambda_2 + 4 \bar \lambda_5 \right)
\left| s \right|^2 s^\ast
+ \left( 12 \bar \lambda_3 + 4 \bar \lambda_4 \right) s^3,
\label{s}
\ee
is the actual \emph{global minimum}\/ of the potential.
A proof that this can actually be achieved is beyond the scope of this paper.

Notice that $\bar \mu$ is supposed to be at the high (seesaw) scale $m_X$,
and then the solution $s$ to eq.~(\ref{s}) will be at that scale too---provided
the coefficients $\bar \lambda_k$ ($k = 1, \ldots, 5$) are of order one.

Finding the minimum of the potential~(\ref{spot}) is a very difficult problem.
However, 
in the special case where $\bar \lambda_3$ and $\bar \lambda_4$ are real
and where $\bar \lambda_2$,
$\bar \lambda_3$,
$\bar \lambda_4$,
and $\bar \lambda_5$ are all \emph{negative},
one can actually prove that $s_1 = s_2 = s_3$
for an adequate range of the parameters of the potential.\footnote{Because
of the invariance of $V_S$ under $S_1 \to - S_1$,
the choice $- s_1 = s_2 = s_3$ will yield an equivalent minimum.
We must assume that Nature has simply chosen $s_1 = s_2 = s_3$
instead of $- s_1 = s_2 = s_3$.}
Indeed,
in that case 
the minimum of $V_S$ with respect to the phases of the VEVs 
will be achieved when all three $s_j$ are real.
One may then write
\be
s_1 = U \cos{\vartheta}, \quad
s_2 = U \sin{\vartheta} \cos{\varphi}, \quad
s_3 = U \sin{\vartheta} \sin{\varphi},
\ee
with $U \ge 0$,
$0 \le \vartheta \le \pi$,
and $0 \le \varphi \le 2 \pi$.
Then,
\be
V_S 
= \bar \mu U^2 + \left( \bar \lambda_1 + \bar \lambda_3 \right) U^4
+ \left( \bar \lambda_2 + \bar \lambda_4 + 2 \bar \lambda_5 \right) U^4
\left( \cos^2{\vartheta} \sin^2{\vartheta}
+ \frac{\sin^4{\vartheta} \sin^2{2 \varphi}}{4} \right).
\ee
If $\bar \lambda_2 + \bar \lambda_4 + 2 \bar \lambda_5 < 0$,
then the minimum will be attained for the values of $\vartheta$
and $\varphi$ that maximize $\cos^2{\vartheta} \sin^2{\vartheta}
+ \left. \left( \sin^4{\vartheta} \sin^2{2 \varphi} \right) \right/ \! 4$.
Assuming $s_j > 0$ for $j=1,2,3$,
these values are $\varphi = \pi / 4$,
$\vartheta = \arccos{1 \left/ \sqrt{3} \right.}$,
corresponding to $s_1 = s_2 = s_3$.

The coefficients $\bar \lambda_3$ and $\bar \lambda_4$
could be real because of an additional $CP$ symmetry. 
That $CP$ symmetry would necessarily
be broken at low scale through the VEVs $v_j$,
which must have different phases
so that the charged-lepton masses are non-degenerate.

In eq.~(\ref{spot}),
the terms with coefficients $\bar \lambda_3$,
$\bar \lambda_4$,
and $\bar \lambda_5$ are sensitive to the phases of the VEVs $s_j$
and \emph{they prevent the emergence of any Goldstone bosons}\/
upon spontaneous symmetry breaking.

However,
one can also take up the opposite stance
and consider the case $\bar \lambda_3 = \bar \lambda_4 = 0$,
which is a special case
of real coefficients $\bar \lambda_3$ and $\bar \lambda_4$.
This may be enforced by a lepton-number ($L$) symmetry under which $D_L$,
$\ell_R$,
$\nu_R$,
and $S$ all carry $L=1$.
This lepton-number symmetry would be broken when the $S_j$ acquire VEVs
and this breaking would lead to a Goldstone boson.
However,
that boson only couples to the right-handed neutrinos---through
the term in eq.~(\ref{LS})---and is,
in practice,
undetectable and harmless~\cite{peccei}.

\end{appendix}

\paragraph{Acknowledgements:}
The work of LL is supported through
the Marie Curie Initial Training Network ``UNILHC'' PITN-GA-2009-237920
and also through the projects PEst-OE-FIS-UI0777-2013,
PTDC/FIS-NUC/0548-2012,
and CERN-FP-123580-2011
of the portuguese {\it Funda\c c\~ao para a Ci\^encia e a Tecnologia}\/ (FCT).

\end{document}